\documentclass[10pt,twocolumn,english,superscriptaddress,noshowpacs]{revtex4}
\usepackage[T1]{fontenc}
\usepackage[utf8]{inputenc}
\usepackage{color}
\usepackage{babel}
\usepackage{amsmath}
\usepackage{amssymb}
\usepackage[unicode=true,
 bookmarks=false,
 breaklinks=false,pdfborder={0 0 1},
 colorlinks=true]
 {hyperref}
\hypersetup{
 pdftex,plainpages=false,citecolor=blue,linkcolor=blue,urlcolor=blue,filecolor=green,bookmarksopen=true}

\makeatletter
\@ifundefined{textcolor}{}
{%
 \definecolor{BLACK}{gray}{0}
 \definecolor{WHITE}{gray}{1}
 \definecolor{RED}{rgb}{1,0,0}
 \definecolor{GREEN}{rgb}{0,1,0}
 \definecolor{BLUE}{rgb}{0,0,1}
 \definecolor{CYAN}{cmyk}{1,0,0,0}
 \definecolor{MAGENTA}{cmyk}{0,1,0,0}
 \definecolor{YELLOW}{cmyk}{0,0,1,0}
}

\usepackage{babel}
\usepackage{graphicx}
\usepackage{epstopdf}\usepackage{amsfonts}
\usepackage{bbm}
\usepackage{color}
\usepackage{latexsym}
\usepackage{times}
\usepackage{txfonts}\usepackage{caption}
\usepackage{slashed}

\def\bs{b\!\!\!/}

\def\Ds{D\!\!\!\!/}

\makeatother

\begin{document}

\title{One-loop calculations in Lorentz-breaking theories and proper-time method}

\author{A. F. Ferrari}
\affiliation{Universidade Federal do ABC, Centro de Ci\^encias Naturais e Humanas, Av. dos Estados, 5001, 09210-580, Santo Andr\'e, SP, Brazil}
\email{alysson.ferrari@ufabc.edu.br}

\author{J. Furtado}
\affiliation{Centro de Ci\^encias e Tecnologia, Universidade Federal do Cariri, 63048-080, Juazeiro do Norte, CE, Brazil}
\email{job.furtado@ufca.edu.br}

\author{J. F. Assun\c c\~ao}
\affiliation{Universidade Regional do Cariri, 63180-000, Juazeiro do Norte, CE, Brazil}
\email{jfassuncao@fis.ufal.br}

\author{T. Mariz}
\affiliation{Instituto de F\'\i sica, Universidade Federal de Alagoas, 57072-900, Macei\'o, AL, Brazil}
\email{tmariz@fis.ufal.br}

\author{A. Yu. Petrov}
\affiliation{Departamento de F\'\i sica, Universidade Federal da Para\'\i ba, Caixa Posta 5008, 58051-970, Jo\~ao Pessoa, PB, Brazil}
\email{petrov@fisica.ufpb.br}

\begin{abstract}
We discuss applications of the proper-time method in various minimal
Lorentz violating modifications of QED and present new results obtained
with its use. Explicitly. we calculate the complete one-loop Heisenberg-Euler
effective action involving all orders in $F_{\mu\nu}$, for two of
the most studied minimal Lorentz-violating extensions of QED, the
one characterized by the axial vector $b^{\mu}$ and the one involving
the second-rank constant tensor $c^{\mu\nu}$.
\end{abstract}

\maketitle

\section{Introduction}

The Fock-Schwinger proper-time method (originally proposed in \cite{Schwinger:1951nm})
presents a powerful way to perform calculations preserving several
formal invariance properties during the calculations. The essence
of this method is based on a representation of the one-loop effective
action as the trace of the logarithm of a field-dependent second-order
differential operator, in the form of an integral involving a function
called the heat kernel. This function turns out to satisfy an ordinary
differential equation which can be solved with a specific technique.
This method has been successfully applied within
various contexts, including gravity (see e.g. \cite{Ojima}) and supersymmetry
(see e.g. \cite{BK}). One of the most convenient versions of the
proper-time method is the one based on the use of the zeta-function
regularization and Gaussian-like integrals \cite{McArthur:1997ww}
which was shown to be efficient for obtaining the Heisenberg-Euler (HE)
action for a scalar QED in a Lorentz-invariant case and its supersymmetric
extension (a general review on the HE effective action
can be found in \cite{Dunne}).

Nowadays, great attention is devoted to studying quantum dynamics
of Lorentz violating (LV) theories, especially to various extensions
of QED (for a review of some basic results obtained within LV theories,
see, e.g., \cite{Colladay:1996iz,Colladay:1998fq,KosPic}). However,
most of these calculations were aimed either to obtaining only contributions
to the quadratic action of external gauge fields (see \cite{Colladay:1998fq,ourED,aether,aether1,Ferr,ourrev,Scarp}
and references therein) or to finding the renormalization of coupling
vertices (see, e.g., \cite{KosPic,Scarp2}). At the same time, the
proper-time method is known to allow for obtaining the complete HE
low-energy effective action whose expression includes all orders in
external gauge fields \cite{BK,McArthur:1997ww}. Therefore it is natural
to expect that the proper-time method can be very useful for explicit
calculating HE-like low-energy effective action in LV
theories.

Up to now, we are aware of a few applications of the proper-time method
within the LV context: the calculation of the non-Abelian Carroll-Field-Jackiw
(CFJ) and the four-dimensional gravitational Chern-Simons terms presented
in\,\cite{ptime}, and also the generation of the Lorentz-breaking
extension of the HE action due to the non-minimal Lorentz-breaking
term $v^{\mu}F_{\mu\nu}\bar{\psi}\gamma^{\nu}\psi$\,\cite{oddptime}.
The relative simplicity of this last case is due to the possibility
of representing the characteristic one-loop determinant of the proper
time method in the simple form $\det(i\Ds-m)$, where $\Ds=\gamma^{\mu}(\partial_{\mu}-ie\tilde{A}_{\mu})$
and $\tilde{A}_{\mu}=A_{\mu}+F_{\mu\nu}v^{\nu}$: this form for the
determinant turns out to be essentially the same as the one appearing
in proper time calculations in Lorentz-invariant cases (see, e.g.,
\cite{McArthur:1997ww,Ojima}). It is interesting to note that not
all cases will allow the rewriting of the determinant in such a familiar
form: if the pseudovector counterpart of this last mentioned model
is considered, for example, the determinant has a much more complicated
form, since the analogue of $\tilde{A}_{\mu}$ will display a
nontrivial matrix structure.

It is known that the pseudovector case, being related to the Adler-Bell-Jackiw
(ABJ) anomaly and the CFJ term, is very compelling from the physical
point of view. Actually, it would be interesting to study the implementation
of the proper-time method for the whole minimal LV extension of QED,
since it is known to be renormalizable, thus one-loop corrections
of all the minimal LV coefficients might, in principle, induce well
defined one-loop corrections. As a first step in this direction, in
this work we will evaluate the one loop effective action for two of
these minimal LV extensions: the one including the $\bs\gamma_{5}$
term is addressed in section II, while the $c^{\mu\nu}\gamma_{\nu}$
term is studied in Section III. In both cases, we will be mainly interested
in the possible contribution of these LV coefficients to the HE
Lagrangian. Finally, Section IV includes our main conclusions.

\section{$\bs\gamma_{5}$ contribution}

This section presents the calculation of one-loop corrections of second
and higher orders in the Lorentz-breaking constant pseudovector $b^{\mu}$.
We start with the model of a fermion coupled to the gauge field via
the Lagrangian 
\begin{equation}
	{\cal L}=\bar{\psi}(i\Ds-m-\bs\gamma_{5})\psi,
\end{equation}
with $D_{\mu}=\partial_{\mu}-ieA_{\mu}$, 
where
the gauge field $A_{\mu}$ can be either Abelian or non-Abelian (with
the inclusion of the relevant trace in this last case). It is interesting
to note that this is probably the most studied part of the minimal
LV extension of the standard model.

The one-loop effective action is given by the fermionic determinant,
\begin{equation}
	\Gamma=i{\rm Tr}\ln(i\Ds-m-\bs\gamma_{5}).\label{ferdet}
\end{equation}
Here, ${\rm Tr}$ stands for the traces over the Dirac matrices as
well as for the integration over the coordinate space. To study the LV extension of the HE action, we must concentrate on the generation of CPT-even terms. In this case
we can sum to the above expression the same term but with inverted
signs of $b_{\mu}$ and the mass, which yields the same result in the CPT-even sector since CPT-even contributions evidently involve only even orders in mass, 
and write 
\begin{equation}
	\Gamma=\frac{i}{2}[{\rm Tr}\ln(i\Ds-m-\bs\gamma_{5})+{\rm Tr}\ln(i\Ds+m+\bs\gamma_{5})],
\end{equation}
which can be rewritten as 
\begin{equation}
	\Gamma=\frac{i}{2}{\rm Tr}\ln[-D^{2}-m^{2}+\frac{ie}{2}\sigma^{\mu\nu}F_{\mu\nu}+2i(b\cdot D)\gamma_{5}-2m\bs\gamma_{5}].\label{square}
\end{equation}
Here we define, unlike the usual convention, $\sigma^{\mu\nu}=\frac{1}{2}[\gamma^{\mu},\gamma^{\nu}]$
and suggest $b_{\mu}$ be light-like, $b^{2}=0$, for simplicity. This
expression can be expanded in power series in $b_{\mu}$, which in principle
can be found order by order.

Let us briefly discuss the possible CPT-odd terms which could arise
from (\ref{ferdet}). Such contributions would involve odd orders
in $b_{\mu}$, hence, their order in the fields would never be equal
to the order in derivatives since the number of all indices is always even. If the number of derivatives in such
a term is more than the number of fields, it involves higher derivatives
and hence must be disregarded as it does not yield the HE
form, and in the opposite situation where the number of derivatives
is less than the number of fields, the resulting term is inconsistent
with the gauge invariance requirement, with the only exception of the CFJ
term. Therefore, it is sufficient to keep only the CPT-even part of
(\ref{ferdet}) for obtaining the HE effective action.

Thus, we can find the contribution of second-order in $b_{\mu}$ from
Eq.~(\ref{square}) to be 
\begin{eqnarray}
	\Gamma_{2} & =&-\frac{i}{4}{\rm Tr}\frac{1}{D^{2}+m^{2}+\frac{1}{2}\sigma^{\mu\nu}F_{\mu\nu}}\times\nonumber\\&\times&
	(2i(b\cdot D)\gamma_{5}-2m\bs\gamma_{5})\times\nonumber \\
	& \times&\frac{1}{D^{2}+m^{2}-\frac{ie}{2}\sigma^{\gamma\delta}F_{\gamma\delta}}(2i(b\cdot D)\gamma_{5}-2m\bs\gamma_{5}).
\end{eqnarray}
This expression can be simplified through the manipulation of the
$\gamma_{5}$ matrices, leading to 
\begin{eqnarray}
	\Gamma_{2} & =&-i{\rm Tr}\frac{1}{D^{2}+m^{2}-\frac{ie}{2}\sigma^{\mu\nu}F_{\mu\nu}}\times\nonumber\\&\times&
	(i(b\cdot D)-m\bs)\nonumber \\
	& \times&\frac{1}{D^{2}+m^{2}-\frac{ie}{2}\sigma^{\gamma\delta}F_{\gamma\delta}}(i(b\cdot D)+m\bs).
\end{eqnarray}
Now, we apply the Schwinger representation 
\begin{equation}
	\frac{1}{D^{2}+m^{2}-\frac{ie}{2}\sigma^{\mu\nu}F_{\mu\nu}}=\int_{0}^{\infty}dse^{-sm^{2}}e^{\frac{ies}{2}\sigma^{\rho\sigma}F_{\rho\sigma}}e^{-sD^{2}},
\end{equation}
for the first fraction, and 
\begin{equation}
	\frac{1}{D^{2}+m^{2}-\frac{ie}{2}\sigma^{\gamma\delta}F_{\gamma\delta}}=\int_{0}^{\infty}dte^{-tm^{2}}e^{\frac{iet}{2}\sigma^{\gamma\delta}F_{\gamma\delta}}e^{-tD^{2}},
\end{equation}
for the other one. We use the constant fields approximation $[D_{\alpha},F_{\beta\gamma}]=0$
as well, just as it was done in \cite{McArthur:1997ww} where this requirement allowed to obtain, in the scalar QED, the all-order result for the one-loop effective action depending only on various degrees of $F_{\alpha\beta}$ but not on its derivatives. In the Abelian case this approximation is equivalent to $\partial_{\alpha}F_{\beta\gamma}=0$. 
So, after considering the cyclic property to put $e^{-sD^{2}}$
to the end of the expression, acting directly on the delta function,
we can write 
\begin{eqnarray}
	\Gamma_{2} & =&-i{\rm tr}\int d^{4}x\int_{0}^{\infty}ds\int_{0}^{\infty}dte^{-(s+t)m^{2}}e^{ie\frac{s}{2}\sigma^{\mu\nu}F_{\mu\nu}}\nonumber \\
	& \times&(i(b\cdot D)-m\bs)e^{ie\frac{t}{2}\sigma^{\rho\sigma}F_{\rho\sigma}}e^{-tD^{2}}(i(b\cdot D)+m\bs)\nonumber \\
	& \times& e^{tD^{2}}e^{-(s+t)D^{2}}\delta^{4}(x-x')|_{x=x'},
\end{eqnarray}
where we inserted the unity in the form $e^{tD^{2}}e^{-tD^{2}}$ for convenience.

Let us now simplify the expression $e^{-tD^{2}}(i(b\cdot D)+m\bs)e^{tD^{2}}$.
While the mass term presents no problems, we must examine $e^{-tD^{2}}D_{\mu}e^{tD^{2}}$.
Since 
\begin{equation}
	e^{-tD^{2}}D_{\mu}e^{tD^{2}}=D_{\mu}-t[D^{2},D_{\mu}]+\frac{t^{2}}{2}[D^{2},[D^{2},D_{\mu}]]+\cdots
\end{equation}
and $[D_{\mu},D_{\nu}]=-ieF_{\mu\nu}$, we then obtain 
\begin{equation}
	e^{-tD^{2}}D_{\mu}e^{tD^{2}}=\exp(-2ietF)_{\mu\nu}D^{\nu}.
\end{equation}
Hence, we can write 
\begin{eqnarray}
	\Gamma_{2} & =&-i\int d^{4}x{\rm tr}\int_{0}^{\infty}ds\int_{0}^{\infty}dte^{-(s+t)m^{2}}e^{\frac{ies}{2}\sigma^{\mu\nu}F_{\mu\nu}}\nonumber \\
	& \times&(i(b\cdot D)-m\bs)e^{\frac{iet}{2}\sigma^{\rho\sigma}F_{\rho\sigma}}
	\times\nonumber\\&\times&
	(i\exp(-2ietF)_{\alpha\beta}b^{\alpha}D^{\beta}+m\bs)\nonumber \\
	& \times& e^{-(s+t)D^{2}}\delta^{4}(x-x')|_{x=x'},
\end{eqnarray}
and, since only even orders in Dirac matrices have a non-zero trace,
we get 
\begin{eqnarray}
	\Gamma_{2} & =&i{\rm tr}\int d^{4}x\int_{0}^{\infty}ds\int_{0}^{\infty}dte^{-(s+t)m^{2}}\nonumber \\
	& \times& \Big(e^{-\frac{(s+t)}{2}\sigma^{\alpha\beta}F_{\alpha\beta}}b_{\lambda}b^{\mu}\exp(-2ietF)_{\mu\nu}D^{\nu}D^{\lambda}\Big.\\
	&+&\Big.m^{2}e^{-\frac{s}{2}\sigma^{\gamma\delta}F_{\gamma\delta}}\bs e^{-\frac{t}{2}\sigma^{\rho\sigma}F_{\rho\sigma}}\bs\Big)\times\nonumber\\&\times&
	e^{-(s+t)D^{2}}\delta^{4}(x-x')|_{x=x'}.\nonumber 
\end{eqnarray}
The expressions $e^{-uD^{2}}\delta^{4}(x-x')|_{x=x'}$ and $D^{\nu}D^{\lambda}e^{-uD^{2}}\delta^{4}(x-x')|_{x=x'}$
have been earlier calculated in \cite{McArthur:1997ww} where they
were shown to yield 
\begin{eqnarray}
	&& D^{\nu}D^{\lambda}e^{-uD^{2}}\delta^{4}(x-x')|_{x=x'}=K^{\nu\lambda}=\nonumber\\&=&
	\left(\frac{-ieF}{e^{-2ieuF}-1}\right)^{\nu\lambda}K(u),\\
	&&e^{-uD^{2}}\delta^{4}(x-x')|_{x=x'}=K(u)=\nonumber\\
	&=&\frac{1}{16\pi^{2}u^{2}}{\rm det}^{1/2}\left(\frac{(ieuF)}{\sinh(ieuF)}\right).
\end{eqnarray}
Thus, after Wick rotation, we can write 
\begin{eqnarray}\label{gamma2}
	\Gamma_{2} & =&\frac{1}{16\pi^{2}}{\rm tr}\int d^{4}x\int_{0}^{\infty}ds\int_{0}^{\infty}dt\frac{1}{(t+s)^{2}}e^{-(s+t)m^{2}}\nonumber \\
	& \times&\Big[e^{ie\frac{(s+t)}{2}\sigma^{\alpha\beta}F_{\alpha\beta}}b_{\lambda}b^{\mu}\times\nonumber\\ &\times&
	\exp(-2ietF)_{\mu\nu}\left(-\frac{ieF}{e^{-2ie(t+s)F}-1}\right)^{\nu\lambda}\Big.\nonumber \\
	& +&\Big.m^{2}e^{ies\sigma^{\rho\sigma}F_{\rho\sigma}}\bs e^{iet\sigma^{\gamma\delta}F_{\gamma\delta}}\bs	\Big]
	\times\nonumber\\&\times&
	{\rm det}^{1/2}\left(\frac{ie(s+t)F}{\sinh(ie(s+t)F)}\right).\;\;
\end{eqnarray}
This expression is of the second order in $b_{\mu}$, but involves arbitrary
orders in $F_{\mu\nu}$. We conclude immediately that this result is gauge
invariant as it must be. The explicit form of the corrections involving
various orders in $F_{\mu\nu}$ can be found through the expansion of
this equation in powers of $F_{\mu\nu}$. We note that this result
is valid both for Abelian and non-Abelian theories. We also note that
higher even orders in $b_{\mu}$ can be obtained in the same manner.

Due to the presence of the matrix traces, the expression (\ref{gamma2}) can be evaluated only order by order. Let us calculate, as an example,
the lower nontrivial (second) order of the expansion of this expression
in $F_{\alpha\beta}$. Explicitly, our aim now consists in finding the lower LV term, that is, the aether-like term.

First, we calculate the traces of products of Dirac matrices. Expanding
the term $e^{\frac{ies}{2}\sigma^{\alpha\beta}F_{\alpha\beta}}\bs e^{\frac{iet}{2}\sigma^{\gamma\delta}F_{\gamma\delta}}\bs$
into power series up to the second order in $F_{\alpha\beta}$, we find 
\begin{eqnarray}
	&&{\rm Tr}\Big(e^{ie\frac{s}{2}\sigma^{\alpha\beta}F_{\alpha\beta}}\bs e^{ie\frac{t}{2}\sigma^{\gamma\delta}F_{\gamma\delta}}\bs\Big)=\nonumber\\
	&=&-8e^2st(bF)^{\alpha}(bF)_{\alpha}+{\rm LI}+\cdots,
\end{eqnarray}
where $(bF)_{\alpha}\equiv b^{\mu}F_{\mu\alpha}$, and ${\rm LI}$ is for Lorentz-invariant
terms proportional to $b^{2}=b^{\mu}b_{\mu}$, which are disregarded within our study since we required $b^2=0$. Here, dots are for fourth-
and higher-order terms in $F_{\alpha\beta}$.

Then, expanding the term $e^{ie(s+t)\sigma^{\alpha\beta}F_{\alpha\beta}}$ in the analogous
power series, we find 
\begin{equation}
	{\rm Tr}e^{ie(s+t)\sigma^{\alpha\beta}F_{\alpha\beta}}=4[1+e^2(s+t)^{2}F^{\alpha\beta}F_{\alpha\beta}]+\cdots.
\end{equation}
This term clearly does not yield a Lorentz-breaking contribution,
so, we disregard the $F_{\alpha\beta}F^{\alpha\beta}$ term since it is irrelevant for our purposes, and write 
\begin{equation}
	{\rm Tr}\, e^{ie(s+t)\sigma^{\alpha\beta}F_{\alpha\beta}}=4[1+e^2(s+t)^{2}F^{\alpha\beta}F_{\alpha\beta}]\simeq4.
\end{equation}
Taking it all together, we arrive at 
\begin{eqnarray}\label{g2}
	\Gamma_{2} & =&\frac{1}{16\pi^{2}}\int d^{4}x\int_{0}^{\infty}ds\int_{0}^{\infty}dt\frac{1}{(t+s)^{2}}e^{-(s+t)m^{2}}\nonumber \\
	& \times&\Big[4b_{\lambda}b^{\mu}\exp(-2ietF)_{\mu\nu}\left(\frac{-ieF}{e^{-2ie(t+s)F}-1}\right)^{\nu\lambda}\Big.+\nonumber\\
	& +&\Big.8(ie)^2m^{2}st(bF)^{\alpha}(bF)_{\alpha}\Big]\times\\ &\times&
	{\rm det}^{1/2}\left(\frac{ie(t+s)F}{\sinh(ie(s+t)F)}\right)+
	O(F^{4}).\nonumber 
\end{eqnarray}

Now, let us keep only the second-order term in $F_{\alpha\beta}$ from this
expression and take into account that (cf. f.e. \cite{oddptime})
\begin{equation}
	{\rm det}^{1/2}\left(\frac{ie(s+t)F}{\sinh(ie(s+t)F)}\right)=1+e^2\frac{(s+t)^{2}}{12}F^{2}+O(F^{4}),
\end{equation}
then, after expanding the exponentials in power series, we obtain
\begin{eqnarray}
	& &\exp(-2ietF)_{\mu\nu}\left(\frac{-ieF}{e^{-2ie(t+s)F}-1}\right)^{\nu\lambda}=\nonumber\\
	&=&	\frac{1}{2(t+s)}(\delta_{\mu}^{\lambda}-ie(t-s)F_{\mu}^{\phantom{\mu}\lambda}\nonumber \\
	& +&\frac{e^2}{3}[t^{2}+s^{2}+8ts](F^{2})_{\mu}^{\phantom{\mu}\lambda})+\cdots.
\end{eqnarray}
Contracting this term with $b_{\lambda}b^{\mu}$, we see that only the term
proportional to $F^{2}$ from this expansion will yield a nontrivial
contribution,
\begin{eqnarray}
	& &\exp(-2ietF)_{\mu\nu}\left(\frac{-ieF}{e^{-2ie(t+s)F}-1}\right)^{\nu\lambda}b^{\mu}b_{\lambda}=\nonumber\\
	&=&-e^2\frac{1}{6(t+s)}[t^{2}+s^{2}+8ts]
	(bF)_{\mu}(bF)^{\mu}+\cdots.
\end{eqnarray}
Substituting all these expansions in (\ref{g2}), we arrive at the following contribution of the second-order
in both $b^{\mu}$ and $F_{\alpha\beta}$: 
\begin{eqnarray}
	\Gamma_{2} & =&-\frac{e^2}{16\pi^{2}}\int d^{4}x\int_{0}^{\infty}ds\int_{0}^{\infty}dt\frac{1}{(t+s)^{2}}e^{-(s+t)m^{2}}\nonumber\\
	& \times&\Big[\frac{2}{3(t+s)}[t^{2}+s^{2}+8ts]\Big.+8m^{2}st\Big](bF)^{\alpha}(bF)_{\alpha}+\nonumber\\&+&
	O(F^{4}).
\end{eqnarray}
Integrating over $t$ and $s$, we obtain 
\begin{eqnarray}
	\Gamma_{2} & = & -\frac{e^2}{6\pi^{2}m^{2}}\int d^{4}x(bF)^{\alpha}(bF)_{\alpha}+
	O(F^{4}).
\end{eqnarray}
Effectively we showed that the lower contribution to the one-loop
effective action, aside from the finite renormalization of the Maxwell
term, is the finite aether term \cite{aether,aether1}, and our result
matches that one from \cite{aether1}, where this contribution was
obtained with the use of Feynman diagrams. This confirms the validity
of our approach. At the same time, while the higher-order terms can
be obtained as well, and they will be evidently finite, their calculation
is more involved since it requires traces of products of a larger
number of Dirac matrices. For example, dimensional reasons restrict the fourth-order result
to be of the form 
\begin{equation}
	\Gamma_{4}=\frac{1}{m^{6}}\int d^{4}x[c_{1}(bF)_{\alpha}(bF)^{\alpha}F^{2}+c_{2}(bF)_{\alpha}F^{\alpha\beta}(bF)^{\gamma}F_{\gamma\beta}],
\end{equation}
where $c_{1},c_{2}$ are dimensionless finite constants. We note that all contributions to the HE involving second and higher orders in $b_{\mu}$ are finite.

It is interesting to compare these results with those ones of the paper \cite{oddptime}. While, within this calculation, unlike  \cite{oddptime}, we deal with a minimal LV extension of QED only, it is interesting to note that, within this calculation, one can make a straightforward replacement $A_{\mu}\to A_{\mu}+F_{\mu\nu}d^{\nu}$, generalizing the results of \cite{oddptime} by a case of presence of two Lorentz-breaking parameters, $b_{\alpha}$ and $d_{\nu}$. Also, we note that while in \cite{aether1}, only lower contributions to the effective action were obtained, our methodology allows to write the complete one-loop low-energy effective action in terms of an unique integral over proper-time parameters.

\section{$c^{\mu\nu}\gamma_{\nu}$ contribution}

In this section, we focus on the contributions for the HE
effective action from the $c^{\mu\nu}\gamma_{\nu}$ Lorentz-breaking term.
Starting from the following Lagrangian, 
\begin{equation}
	\mathcal{L}=\bar{\psi}(i\gamma^{\mu}D_{\mu}-m+iD_{\mu}c^{\mu\nu}\gamma_{\nu})\psi,
\end{equation}
one can obtain the effective action $\Gamma$ by the fermion integration,
which yields 
\begin{eqnarray}
	\Gamma & =&-iTr\ln(iD_{\mu}\gamma^{\mu}-m)-\nonumber \\
	& -&iTr\ln\left(1+\frac{1}{iD_{\mu}\gamma^{\mu}-m}iD_{\lambda}c^{\lambda\nu}\gamma_{\nu}\right).
\end{eqnarray}
Thus, the effective action can be presented as the sum of two contributions,
where the second one will be written in an integral form, as follows,
\begin{equation}
	\Gamma=S^{(0)}+S^{(1)},
\end{equation}
with 
\begin{eqnarray}
	S^{(0)} & =&-iTr\ln(iD_{\mu}\gamma^{\mu}-m),\\
	S^{(1)} & =&-i\int_{0}^{1}dzTr\left[\frac{1}{-iD_{\mu}\gamma^{\mu}+m-izD_{\lambda}c^{\lambda\nu}\gamma_{\nu}}\right.\nonumber \\
	& \times& \left.(-iD_{\sigma}c^{\sigma\rho}\gamma_{\rho})\right].
\end{eqnarray}
The contribution $S^{(0)}$ is the usual QED contribution while $S^{(1)}$
includes the Lorentz-breaking term. We note that in this case, the
LV parameter $c^{\mu\nu}$ is not accompanied by any Dirac matrix, so
all trace calculations are much simpler, and we will not need to calculate
order by order in $c_{\mu\nu}$, instead finding a closed form result.
As we are interested in the Lorentz-violating effects of the tensor
$c^{\mu\nu}$ to the HE effective action, we will work only
with the contribution $S^{(1)}$.

At this point, following Schwinger's procedure \cite{Schwinger:1951nm},
we introduce the fermionic Green's function 
\begin{equation}
	(-iD_{\mu}\gamma^{\mu}+m-izD_{\lambda}c^{\lambda\nu}\gamma_{\nu})G(x,x')=\delta^{4}(x-x').\label{fermgreenfunc}
\end{equation}
Consequently, $S^{(1)}$ can now be rewritten as 
\begin{equation}
	S^{(1)}=-i\int d^{4}x\int_{0}^{1}dz\thinspace{\rm tr}[G(x,x')(-iD_{\lambda}c^{\lambda\mu}\gamma_{\mu})]|_{x\rightarrow x'}.
\end{equation}
Here, tr means the trace only over the spinor indices. Defining the
bosonic Green's function as 
\begin{equation}
	G(x,x')=(iD_{\mu}\gamma^{\mu}+izD_{\lambda}c^{\lambda\nu}\gamma_{\nu}+m)\Delta(x,x'),\label{bosgreenfunc}
\end{equation}
after we replace it into (\ref{fermgreenfunc}), we obtain the following
equation for it: 
\begin{eqnarray}
	& &\left[D^{2}+m^{2}-\frac{ie}{2}\sigma_{\mu\nu}F^{\mu\nu}+D_{\mu}\gamma^{\mu}D_{\beta}c^{\beta\delta}\gamma_{\delta}+\right.\nonumber\\ &+&\left.zD_{\alpha}c^{\alpha\nu}\gamma_{\nu}D_{\mu}\gamma^{\mu}\right.\\
	 &-&\left.z^{2}(D_{\alpha}c^{\alpha\nu}\gamma_{\nu})(D_{\beta}c^{\beta\delta}\gamma_{\delta})\right]\Delta(x,x')=\delta^{4}(x-x').\nonumber
\end{eqnarray}
Therefore, the above equation can be written as 
\begin{equation}
	\mathcal{H}\Delta(x,x')=\delta^{4}(x-x'),\label{H1}
\end{equation}
where $\mathcal{H}$, 
\begin{eqnarray}
	\mathcal{H} & =&D^{2}+m^{2}-ie\frac{1}{2}\sigma_{\mu\nu}F^{\mu\nu}+2zD_{\delta}D_{\beta}c^{\beta\delta}-\nonumber\\
	&-& iezF_{\mu\beta}c^{\beta\delta}\gamma_{\delta}\gamma^{\mu}\nonumber \\
	& -&z^{2}D_{\alpha}D_{\beta}c^{\alpha\nu}c^{\beta\delta}g_{\nu\delta}+\frac{ie}{2}z^{2}F_{\alpha\beta}c^{\beta\delta}c^{\alpha\nu}\gamma_{\delta}\gamma_{\nu},\label{H2}
\end{eqnarray}
is identified as the Hamiltonian. The idea of identifying $\mathcal{H}$
in (\ref{H1}) and (\ref{H2}) as the Hamiltonian of a hypothetical
quantum mechanical system, whose evolution is given by the \textit{\emph{time}}
parameter $s$, is one of the central points of the Fock-Schwinger
proper-time method. However, due to the smallness of the Lorentz violating
tensor $c^{\mu\nu}$, we will consider only the first-order contributions
in $c^{\mu\nu}$, or, as is the same in our case, in the parameter $z$, 
to reduce the Hamiltonian to 
\begin{equation}
	\mathcal{H}=D^{2}+m^{2}-\frac{ie}{2}\sigma_{\mu\nu}F^{\mu\nu}+2zD_{\delta}D_{\beta}c^{\beta\delta}-iezF_{\mu\beta}c^{\beta\delta}\gamma_{\delta}\gamma^{\mu}.
\end{equation}

In order to obtain the Hamiltonian in the Heisenberg representation,
one first has to know the evolution of $x(s)$ and $D(s)$ in terms
of the proper time parameter $s$. To achieve such evolution one must
use the standard commutations relations, 
\begin{equation}
	[x_{\mu},D_{\nu}]=-g_{\mu\nu},\,[D_{\mu},D_{\nu}]=-ieF_{\mu\nu},
\end{equation}
and, consequently, the equations of motion for the operators $x(s)$
and $D(s)$ are: 
\begin{eqnarray*}
	\dot{x}_{\mu} & =&i[\mathcal{H},x_{\mu}]=2iD_{\mu}+4izc_{\mu}^{\phantom{\mu}\nu}D_{\nu},\\
	\dot{D}_{\mu} & =&i[\mathcal{H},D_{\mu}]=-2eF_{\mu}^{\phantom{\mu}\nu}D_{\nu}-4ezF_{\mu}^{\phantom{\mu}\alpha}c_{\alpha}^{\phantom{\alpha}\beta}D_{\beta},
\end{eqnarray*}
where we again assumed $F_{\alpha\beta}$ to be constant
Rewriting the previous equations in
matricial notation we arrive at 
\begin{eqnarray}
	x(s) & =&x(0)+\Lambda^{-1}BiD(0)\left(e^{\Lambda s}-1\right),\label{x}\\
	D(s) & =&D(0)e^{\Lambda s},\label{pi}
\end{eqnarray}
where 
\begin{eqnarray}
	\Lambda & =&2eF+4eFc,\\
	B & =&2+4zc.
\end{eqnarray}
The equations (\ref{x}) and (\ref{pi}) were obtained by choosing, for the sake of the convenience, the partricular form of the tensor $c^{\mu\nu}$ so that it has the following components:
\begin{equation}
	\label{kappa}
	c^{ij}=0\,\, {\rm at}\,\, i\neq j;\quad\,  c^{00}=c^{ii}=\kappa. 
\end{equation}
In this case the matrices
$\Lambda^{-1}$ and $B$ obey the commutation relation $[\Lambda^{-1},B]=0$,
besides, $\Lambda$ is anti-symmetric and therefore traceless.
The condition imposed on the LV tensor $c^{\mu\nu}$ is the more convenient
choice in order to guarantee the validity of the method in the present
context. From the relations (\ref{x}) and (\ref{pi}), one can find
\begin{eqnarray}
	D(0) & =&-\frac{i}{2}\Lambda B^{-1}e^{-\frac{1}{2}\Lambda s}\sinh^{-1}\left(\frac{1}{2}\Lambda s\right)\left[x(s)-x(0)\right]\nonumber \\
	& =&-\frac{i}{2}iFe^{-\frac{1}{2}\Lambda s}\sinh^{-1}\left(\frac{1}{2}\Lambda s\right)\left[x(s)-x(0)\right],\\
	D(s) & =&-\frac{i}{2}\Lambda B^{-1}e^{\frac{1}{2}\Lambda s}\sinh^{-1}\left(\frac{1}{2}\Lambda s\right)\left[x(s)-x(0)\right]\nonumber \\
	& =&-\frac{i}{2}eFe^{\frac{1}{2}\Lambda s}\sinh^{-1}\left(\frac{1}{2}\Lambda s\right)\left[x(s)-x(0)\right],\\
	D^{2}(s) & =&-[x(s)-x(0)]K[x(s)-x(0)],
\end{eqnarray}
with $K$ being defined as 
\begin{equation}
	K=\frac{1}{4}\left(eF\right)^{2}\sinh^{-2}\left(\frac{1}{2}\Lambda s\right).
\end{equation}
Note that  $D^2(s)$ can be written as
\begin{eqnarray}
	D^2(s) & =&x(s)Kx(s)-2x(s)Kx(0)-x(0)Kx(s)\nonumber \\
	& +&x(s)Kx(0)+x(0)Kx(0).
\end{eqnarray}
Using the following commutation relation 
\begin{equation}
	[x_{\mu}(s),x_{\nu}(0)]=-\left[\left(-ieF\right)^{-1}(e^{\Lambda s}-1)\right]_{\mu\nu},
\end{equation}
it is possible to find that 
\begin{equation}
	x(s)Kx(0)-x(0)Kx(s)=-\frac{1}{2}TR\left[F\coth\left(\frac{1}{2}\Lambda s\right)\right].
\end{equation}
In the above equation the trace $TR$ stands only over the Lorentz
indices, i.e., $TR(AB)=A_{\mu\nu}B^{\nu\mu}$. Therefore the Hamiltonian
becomes 
\begin{eqnarray}
	\mathcal{H} & =&-\Delta x\frac{BK}{2}\Delta x-\frac{i}{4}TR\left[\Lambda\coth\left(\frac{1}{2}\Lambda s\right)\right]\nonumber \\
	& -&\frac{1}{4}TR\left(\sigma\Lambda\right)+m^{2},
\end{eqnarray}
where $\Delta x=x'-x''=x(s)-x(0)$. However, because of the relation
$i\partial_{s}\langle x'(s)|x''(0)\rangle=\langle x'(s)|\mathcal{H}|x''(0)\rangle$,
we have that
\begin{eqnarray}
	i\partial_{s}\langle x'(s)|x''(0)\rangle & =&\left\{ -\Delta x\frac{BK}{2}\Delta x-\right.\nonumber\\
	&-&\left.
	\frac{i}{4}TR\left[\Lambda\coth\left(\frac{\Lambda}{2}s\right)\right]\right. \\
 &-&\left.\frac{1}{4}TR\left(\sigma\Lambda\right)+m^{2}\right\} \langle x'(s)|x''(0)\rangle,\nonumber
\end{eqnarray}
whose solution for $\langle x'(s)|x''(0)\rangle$ can be written as
\begin{eqnarray}\label{ProdInt}
	\langle x'(s)|x''(0)\rangle & =&C(x',x'')s^{-2}e^{-L(s)}\times\nonumber \\
	& \times&\exp\left[-\frac{i}{8}\Delta x\Lambda\coth\left(\frac{1}{2}\Lambda s\right)\Delta x\right.\nonumber \\
	 &-&\left.\frac{1}{4}TR\left(\sigma\Lambda\right)-im^{2}s\right],
\end{eqnarray}
where 
\begin{equation}
	L(s)=\frac{1}{2}TR\ln\left[\left(\frac{\Lambda}{2}s\right)^{-1}\sinh\left(\frac{1}{2}\Lambda s\right)\right].
\end{equation}
The function $C(x',x'')$ was determined by Schwinger in \cite{Schwinger:1951nm}
as being 
\begin{eqnarray}
	C(x',x'')=C\Phi(x',x''),
\end{eqnarray}
with 
\begin{eqnarray}
	\Phi(x',x'')=\exp\left[ie\int_{x''}^{x'}dxA(x)\right]
\end{eqnarray}
and $C$ is the constant $C=-i(4\pi)^{-2}$. The functional similarity
between $\langle x'(s)|x''(0)\rangle$ in (\ref{ProdInt}) and the $C(x',x'')$ found in \cite{Schwinger:1951nm},
as well as the properties of $\Lambda$ which are the same as of $F$,
guarantees that the same result for $C$ can be applied for our case.
Following the proper-time procedure, the Lagrange function can be
now found from the expression 
\begin{equation}
	\mathcal{L}^{(1)}(x)=\frac{i}{2}\int_{0}^{1}dz\int_{0}^{\infty}dss^{-1}tr\langle x'(s)|x''(0)\rangle|_{x',x''\rightarrow x},
\end{equation}
so, after the substitution of the result of $\langle x'(s)|x''(0)\rangle$
we obtain 
\begin{eqnarray}
	\mathcal{L}^{(1)}(x) & = & \frac{1}{32\pi^{2}}\int_{0}^{1}dz\int_{0}^{\infty}dss^{-3}\exp{(-im^{2}s)}e^{-L(s)}\times\nonumber \\
	&  & \times tr\left[\exp\left(\frac{i}{4}TR(\sigma\Lambda)s\right)\right].
\end{eqnarray}
Deforming the integration path to the positive real axis by performing
the Wick rotation $s\rightarrow-is$, we arrive at 
\begin{eqnarray}
	\mathcal{L}^{(1)}(x) & =&-\frac{1}{32\pi^{2}}\int_{0}^{1}dz\int_{0}^{\infty}dss^{-3}\exp{(-m^{2}s)}e^{-l(s)}\times\nonumber \\
	& \times& tr\left[\exp\left(\frac{1}{4}TR(\sigma\Lambda)s\right)\right],
\end{eqnarray}
with $l(s)$ defined as
\begin{equation}
	l(s)=\frac{1}{2}TR\ln\left[\left(\frac{\Lambda}{2}s\right)^{-1}\sin\left(\frac{1}{2}\Lambda s\right)\right].
\end{equation}
The properties of $\Lambda$ allow us to use the eigenvalue technique
\cite{Schwinger:1951nm}, so that $e^{-l(s)}$ turns out to be 
\begin{equation}
	e^{-l(s)}=\frac{(es)^{2}\mathcal{Q}}{Im[\cosh(esY)]},
\end{equation}
with 
	\begin{eqnarray}
		Y&=&\sqrt{2(\mathcal{T}+i\mathcal{Q})},\\
		\mathcal{T}&=&\mathcal{F}-e^2zF_{\mu\nu}F^{\mu\alpha}c_{\alpha}^{\,\,\,\,\nu}=\nonumber\\
		&=&-\frac{e^2}{2}(\vec{E}^2-\vec{B}^2)(1+4z\kappa),\\
		\mathcal{Q}&=&\mathcal{G}-e^2z\kappa F_{\mu\nu}\tilde{F}^{\mu\nu}=e^2\vec{E}\cdot\vec{B}(1+4z\kappa),\\
		\mathcal{F}&=&-\frac{e^2}{4}F_{\mu\nu}F^{\mu\nu}=-\frac{e^2}{2}(\vec{E}^2-\vec{B}^2),\\
		\mathcal{G}&=&\frac{1}{4}F_{\mu\nu}\tilde{F}^{\mu\nu}=-\vec{E}\cdot\vec{B},
	\end{eqnarray}
where $\kappa$ is defined by (\ref{kappa}).

Therefore, the final result for $\mathcal{L}^{(1)}(x)$ is 
\begin{eqnarray}
	\mathcal{L}^{(1)}(x) & =&-\frac{1}{8\pi^{2}}\int_{0}^{1}dz\int_{0}^{\infty}s^{-3}\exp(-m^{2}s)\times\nonumber \\
	& \times&\left[(es)^{2}\mathcal{Q}\frac{Re[\cosh(esY)]}{Im[\cosh(esY)]}-1\right].
\end{eqnarray}
The additive constant $-1$ was added, following Schwinger's procedure,
in order to guarantee that the Lagrangian will vanish when all fields
are set to zero. Now, we remember the quadratic part of the Lagrangian
of the Maxwell field in the presence of Lorentz symmetry breaking,
\begin{equation}
	\mathcal{L}^{(0)}=\mathcal{T}=-\frac{e^2}{2}(\vec{E}^{2}-\vec{B}^{2})(1+4\kappa).
\end{equation}
and including this in our result, we obtain the finite gauge invariant
resulting Lagrangian as
\begin{eqnarray}
	\mathcal{L} & =&\mathcal{T}-\int_{0}^{1}dz\int_{0}^{\infty}dss^{-3}e^{-m^{2}s}\nonumber \\
	& \times&\left[(es)^{2}\mathcal{Q}\frac{Re[\cosh(esY)]}{Im[\cosh(esY)]}-1-\frac{2}{3}(es)^{2}\mathcal{T}\right],
\end{eqnarray}
which can be rewritten explicitly in terms of the electric and magnetic
fields as 
\begin{eqnarray}
	\mathcal{L} & =&-\frac{e^2}{2}(\vec{E}^{2}-\vec{B}^{2})(1+4\kappa)+\nonumber\\&+&
	\frac{2\alpha^{2}\hbar^{3}}{45m^{4}c^{5}}\left[(\vec{E}^{2}-\vec{B}^{2})^{2}(1+4\kappa)^{2}+\right.\nonumber \\
	&+& \left.7(\vec{E}\cdot\vec{B})^{2}(1+4\kappa)^{2}\right]+\cdots.
\end{eqnarray}
The above Lagrangian is the HE action corrected by the
LV contribution proportional to $c^{\mu\nu}$. The result resembles the
usual QED expression for the HE action, which confirms a result previously
found in \cite{FM} for the HE action by the explicit calculation
of the Feynman diagrams. We can also note that the contribution $\mathcal{T}$
correspond to the one-loop correction to the Maxwell Lagrangian given
by the LV $c^{\mu\nu}$ term which was already presented in \cite{Mariz:2016ooa}.
We note that the methodology we presented here allows for an automatic summation over all orders in external gauge fields.
Another way of treating the contribution involving the $c^{\mu\nu}$ consists in its reabsorbing
it into the effective metric (see e.g. \cite{ColMac,Scarpelli:2015iia}).

\section{Final Remarks}

In this paper, we presented some applications of the proper-time method in Lorentz-breaking
modifications of QED. We started with the LV additive term $\bar{\psi}\bs\gamma_{5}\psi$
and we applied the proper-time method to find the one-loop contribution
to the electromagnetic effective action involving all orders in a
constant external $F_{\mu\nu}$, which corresponds to constant external electric
and magnetic fields. We argued that the only CPT-odd contribution
arising in this case, without higher derivatives, is the CFJ term.

We also studied the generation of the HE effective action
in the CPT-even sector of the minimal Lorentz-violating extension
of the standard model proposed by Kostelecky \cite{Colladay:1996iz,Colladay:1998fq},
more specifically from the term proportional to $c^{\mu\nu}$. Considering
only contributions linear in the LV coefficient $c^{\mu\nu}$ we found
a non-linear correction to the QED Lagrangian that resembles the usual
HE correction. The result we found matches the previous
works on quadratic radiative corrections \cite{Mariz:2016ooa} and on HE effective
action \cite{FM}.

We close the text with the discussion of the possibility of using
the proper-time method for other minimal Lorentz-breaking extensions
of QED \cite{KosPic}. The generic minimal LV extension of the spinor
part of the QED Lagrangian defined in this paper reads
\begin{equation}
	{\cal L}=\bar{\psi}(i\Gamma^{\nu}D_{\nu}-M)\psi,\label{genrenmod}
\end{equation}
where 
\begin{eqnarray}
	\Gamma^{\nu} &=&\gamma^{\nu}+c^{\mu\nu}\gamma_{\mu}+d^{\mu\nu}\gamma_{\mu}\gamma_{5}+e^{\nu}+if^{\nu}\gamma_{5}+\nonumber\\
	&+&\frac{1}{2}g^{\lambda\mu\nu}\sigma_{\lambda\mu},\nonumber \\
	M &=&m+a_{\mu}\gamma^{\mu}+b_{\mu}\gamma^{\mu}\gamma_{5}+\frac{1}{2}H^{\mu\nu}\sigma_{\mu\nu}.
\end{eqnarray}
In this paper, we discussed the contributions originated on $b^{\mu}$
and $c^{\mu\nu}$. The contribution of $a^{\mu}$ can be ruled out from
the action with use of a gauge transformation. The calculations of
the analogues of the HE effective action generated by $d^{\mu\nu}$, $e^{\mu}$, $f^{\mu}$, $g^{l\mu\nu}$ and
$H^{\mu\nu}$ within the proper-time framework still represent open problems.

{\bf Acknowledgments.}
This study was partially supported by Conselho Nacional de Desenvolvimento Cient\'\i fico e Tecnol\'ogico (CNPq) via the grants 301562/2019-9 (AYP), 310066/2018-2 (TM), and 305967/2020-7 (AFF).

\end{document}